\documentclass[aip,reprint]{revtex4-1}
\usepackage[hidelinks,colorlinks=true,citecolor=blue]{hyperref}
\usepackage{amsmath}    % 必须优先加载
\usepackage{braket}     % 符号宏包在后
\usepackage{graphicx}% Include figure files
\usepackage{dcolumn}% Align table columns on decimal point
\usepackage{bm}% bold math

\usepackage[utf8]{inputenc}
\usepackage[T1]{fontenc}
\usepackage{mathptmx}
\usepackage{etoolbox}

\makeatletter
\def\@email#1#2{%
 \endgroup
 \patchcmd{\titleblock@produce}
  {\frontmatter@RRAPformat}
  {\frontmatter@RRAPformat{\produce@RRAP{*#1\href{mailto:#2}{#2}}}\frontmatter@RRAPformat}
  {}{}
}%
\makeatother
\begin{document}

\preprint{}

\title{High-sensitivity and high-resolution collaborative determination of birefringence coefficient using weak measurement}

\author{Shuqi Gao}%
\author{Min Zhang}%
\affiliation{Key Laboratory of Advanced Transducers and Intelligent Control System, Ministry of Education, Taiyuan University of Technology, Taiyuan 030024, China}%
\affiliation{College of Physics and Optoelectronics, Taiyuan University of Technology, Taiyuan 030024, China}%

\author{Jiahui Hou}%
\affiliation{Key Laboratory of Advanced Transducers and Intelligent Control System, Ministry of Education, Taiyuan University of Technology, Taiyuan 030024, China}%
\affiliation{College of Physics and Optoelectronics, Taiyuan University of Technology, Taiyuan 030024, China}%
\affiliation{Shanxi Key Laboratory of Precision Measurement Physics, Taiyuan University of Technology, Taiyuan 030024, China}%

\author{Qingchen Liu}%
\affiliation{Key Laboratory of Advanced Transducers and Intelligent Control System, Ministry of Education, Taiyuan University of Technology, Taiyuan 030024, China}%
\affiliation{College of Physics and Optoelectronics, Taiyuan University of Technology, Taiyuan 030024, China}%

\author{Hongyu Li}%
\affiliation{College of Physics and Optoelectronics, Taiyuan University of Technology, Taiyuan 030024, China}%
\affiliation{Shanxi Key Laboratory of Precision Measurement Physics, Taiyuan University of Technology, Taiyuan 030024, China}%

\author{Xiaomin Guo}%
\affiliation{Key Laboratory of Advanced Transducers and Intelligent Control System, Ministry of Education, Taiyuan University of Technology, Taiyuan 030024, China}%
\affiliation{College of Physics and Optoelectronics, Taiyuan University of Technology, Taiyuan 030024, China}%

\author{Yanqiang Guo}%
\altaffiliation{Electronic mail: guoyanqiang@tyut.edu.cn}%Lines break 
\affiliation{Key Laboratory of Advanced Transducers and Intelligent Control System, Ministry of Education, Taiyuan University of Technology, Taiyuan 030024, China}%

\affiliation{College of Physics and Optoelectronics, Taiyuan University of Technology, Taiyuan 030024, China}%

\affiliation{Shanxi Key Laboratory of Precision Measurement Physics, Taiyuan University of Technology, Taiyuan 030024, China}%

\author{Liantuan Xiao}%
\affiliation{Key Laboratory of Advanced Transducers and Intelligent Control System, Ministry of Education, Taiyuan University of Technology, Taiyuan 030024, China}%
\affiliation{College of Physics and Optoelectronics, Taiyuan University of Technology, Taiyuan 030024, China}%

\begin{abstract}
Precise nanofilm birefringence characterization is essential for high-sensitivity polarization response and strong anti-interference detection in photodetectors. We present a high-sensitivity and high-resolution birefringence coefficient determination system for nm-level membranes based on weak measurement, addressing the sensitivity-resolution trade-off. A tunable bandwidth light source is exploited to achieve simultaneous and complementary measurements of momentum (P-pointer) and intensity (I-pointer), enabling calibration-free operation across various bandwidths, and to realize high-precision phase difference monitoring of the measured membranes. This method maps the birefringence effect to a weak-value amplified signal of spectral shift and light intensity. The optimal resolution, achieved at a spectral width of 6 nm, is $1.12 \times 10^{-8}$ RIU, while the optimal sensitivity is achieved when the light source is a narrow-linewidth coherent laser, reaching 4710 mV/RIU. The linear range of the system covers a broad birefringence coefficient range for crystals, from $10^{-6}$ to 0.1. Furthermore, the auxiliary optical path eliminates substrate interference, achieving a detection limit of birefringence coefficient as low as $10^{-8}$ RIU. This approach, characterized by high precision, high sensitivity, and strong robustness, provides an effective solution for the detection of optical nano-thin membrane parameters.
\end{abstract}

\maketitle

Birefringence coefficient is a key physical parameter of optical membranes \cite{Jia25,Mas25} and directly influences the polarization characteristics \cite{Xav24}, transmission efficiency and stability of the corresponding optical devices. And it is one of the most concerning indicators when it refers to the designing and optimizing of optical films \cite{Kur25}. For high-performance photodetection in extreme environments, precise nanofilm birefringence characterization underlies the optimization of polarization response sensitivity and signal-to-noise ratio, and critically supports the improvement of device stability, detection accuracy, and anti-interference performance in complex optical fields. To date, the birefringence coefficient can be measured by polarization optical instruments and optical interferometers \cite{Das24}. However, results from polarization optical instruments are extremely susceptible to thickness and surface quality of samples, which restricts the measurement accuracy. Although optical interferometers can achieve high-accuracy determination of the birefringence coefficient under carefully controlled conditions, the operational complexity and the rigorous test environment requirements still limit their applications \cite{Sun25}. Moreover, most of the existing measurement technologies generally face issues such as limited coherence of light sources and low measurement efficiency, making it difficult to meet the practical demands for high precision and effective detection \cite{KIM22}. This is especially true in scenarios like micro/nano thin membrane property analysis or complex structure processing, where the limitations of traditional methods are further highlighted. The rigorous requirements for sample surface morphology, environmental stability, or dynamic processes often lead to a decrease in the reliability of measurement results. Therefore, there is an urgent need to develop more efficient and versatile measurements to overcome the existing barriers in precision and sensitivity.

In recent years, quantum weak measurement, with its high sensitivity and anti-noise capabilities \cite{luo1,zhang1,luo3}, has provided a new approaches for detecting small physical quantities \cite{luo2,zhang3}, which can be applied to develop birefringence coefficient measurement technology \cite{Jo07,You18}. The measurement uses the weak value amplification effect \cite{he1,guo1,he2} to transform the undetectable measurement-targeted parameters into observable light intensity or spectral shift signals to significantly improving measurement resolution \cite{JS11,Fe14,zhang2,Li11}. Currently, weak measurement is constrained by the spectral width fluctuations of light sources and the noise of system \cite{zhang6,luo2}, which restricts measurement accuracy and dynamic range \cite{Q17,Xu20,Luo19}. For instance, narrow-spectral light sources lack sufficient sensitivity for light intensity indicators \cite{guo2,Peng,Bru10,Xu13}, while wide-spectral light sources are prone to momentum spectral broadening errors \cite{he3,Hu19,guo2024,zhang5,Zhang4}, thereby making it difficult to meet robustness requirements across various scenarios.

To address these challenges, this work presents a high-sensitivity and high-resolution nanometer-level birefringence coefficient detection system for thin membranes based on weak measurement. The complementary momentum P and intensity I dual-parameter measurements are designed, successfully breaking the sensitivity‑resolution trade‑off. The measurements are also combined with shared optical paths that require no recalibration across different bandwidths, and phase difference detection to constitute the thin membrane birefringence coefficient measurement system. The system synchronously maps the birefringence effect into weak amplified signals of spectral shift and light intensity variation. We achieve optimized resolution for both the P-pointer and I-pointer across a range of spectral conditions, from narrow to wide bandwidths. The resolution of the P-pointer is on the order of $10^{-8}$ RIU, significantly higher than the $10^{-4}$ RIU resolution of ellipsometry. In addition to maintaining high resolution, the system eliminates substrate interference through an auxiliary optical path, achieving highly linear detection of the birefringence coefficient, with the I-pointer sensitivity reaching above $4.7 \times 10^{3}$ mV/RIU exceeding current surface plasmon resonance detection methods. Its simultaneous measurement mechanism demonstrates the complementary advantages of variable bandwidth, with a wider spectrum optimizing the dynamic range and noise immunity of the P-pointer, and a narrower spectrum enhancing the sensitivity of the I-pointer. Furthermore, by introducing high-gain detection and dynamic noise suppression strategies, the system maintains a high and robust signal-to-noise ratio in complex noise environments. The measurement system exhibits the advantages of high resolution, high sensitivity, operational simplicity, and strong robustness, which provides a useful solution for the precise measurement of optical thin membrane parameters.

We first introduce an optimized shared optical path weak measurement system, which is used to measure small phase differences $\varphi$. The initial polarization of the system is $|\psi_{i}\rangle = (1/\sqrt{2})(|H\rangle + |V\rangle)$, and the momentum spectrum distribution of the light source is represented as $|\phi(p)\rangle$. The photon momentum is $p_{0} = \omega_{0}/c$, where $\omega_{0}$ represents the central frequency corresponding to the center wavelength $\lambda_{0}$ of the incident light , and $c$ is the speed of light. 

In weak interaction, one of the HWPs (Half-Wave Plates) is tilted by a small angle of $\theta$ to introduce an extremely small net phase shift of $\Delta$ between light $o$ and light $e$. The mapping relationship is $\Delta  = \pi\bigl(1/\sqrt{1 - \sin^2\!\theta/n^2} - 1\bigr)$, where $n=1.54$ is the refractive index corresponding to the true zero-order HWP at 1550 nm in the experiment.the first-order approximation model demonstrates nearly identical wavelength shifts across weak interactions, thereby enabling precise birefringence coefficient measurement. The quantitative correspondence between birefringence coefficient and phase shift is derived as:
  \begin{equation}
  \Delta\varphi = \frac{2\pi}{\lambda} \cdot \Delta n \cdot d \cdot \frac{1}{\cos\theta}. 
  \label{eq1}
\end{equation}
Here, $\Delta=\Delta\varphi$, $\lambda=1550nm$, and $\Delta n$ represent the birefringence coefficients of the test sample that this paper focuses on, $d$ is the film thickness, and $1 / \cos\theta$ reflects the modulation effect of the tilt angle on the optical path difference. Additionally, an auxiliary optical path is established as a monitoring path to ensure $\theta_{sample}=0$. 

The P-pointer was analyzed. The meter state is initialized to $\int \! \mathrm{d}p \, \ket{\phi(p)}$, which is a Gaussian function with center $p_0$ and standard deviation $\sigma_{p}$. The interaction strength $g = \Delta n \cdot d$, which is proportional to the phase difference $\Delta\varphi$ with a coefficient depending on the tilt angle $\theta$, can be derived from the weak interaction between the target quantum system and the meter state. For the weak interaction, the interaction operator is represented as $\hat{U} = \exp\!\bigl(ig/2 \hat{A} \otimes \hat{P}\bigr)$, where $\hat{A} = |H\rangle\langle H| - |V\rangle\langle V|$ is the observable of the target system, and $\hat{P}$ is the momentum operator of the meter state. Post-selection is subsequently performed on the system within the selected projection basis, typically projected onto a state that is nearly orthogonal to the initialization $|\psi_{f}\rangle = 1/\sqrt{2} \bigl( e^{i\rho}\!\ket{H} - e^{i\rho}\!\ket{V} \bigr)$, where $\rho$ is the post-selection angle. Post-selection induces the collapse of the measurement instrument state into an unnormalized redistribution as $D(p) = \Phi(p)/2 \bigl[ 1 - \cos\!\left(gp + 2\rho\right) \bigr]$, $\Phi(p) = \left| \langle \phi(p) | \phi(p) \rangle \right|^2$ is the spatial distribution of the initial momentum and $p$ is the eigenvalue of $\hat{P}$. When $\rho \leq 1$, the post-selection causes the measurement instrument state to collapse into a bimodal distribution. The observable post-selection probability is $P \approx \sin^{2}(\rho)$, and the weak value $\langle \hat{A} \rangle_w$ can be obtained:
 \begin{equation}
 \langle\hat{A}\rangle_w = \frac{\langle\psi_f|\hat{A}|\psi_i\rangle}{\langle\psi_f|\psi_i\rangle} = i\cot(\rho). 
  \label{eq2}
\end{equation}

It is noteworthy that the interaction strength $g$ can be extracted from the offset of the average value of the P-pointer, and the expression is:
\begin{equation}
 \begin{split}
\Delta p &= \frac{\int pD(p)\,\mathrm{d}p}{\int D(p)\,\mathrm{d}p} - p_0 \\
&= \frac{1}{2P}\sigma_p^2 g\, e^{-\sigma_p^2g^2} \sin\!\left(gp_0 + 2\rho\right) \approx g\sigma_p^2\,\mathrm{Im}\!\left(\langle\hat{A}\rangle_w\right).
\end{split}
  \label{eq3}
\end{equation}

The weak measurement approximately satisfies $gp_0/2 \ll \rho \ll 1$ within the entire linear range of the experiment. Given that the light intensity I-pointer represents intensity quantity, the initial light intensity without post-selection is denoted by $I_{g=0}$. After implementing weak interaction and post-selection $|\psi_{f}\rangle$, the light intensity received by the detector evolves to $I= I_{\text{g=0}} P$. The shift of I-pointer is: 
\begin{equation}
\Delta L = \frac{\Delta I}{I_{g=0}} = \frac{I - I_{g=0}}{I_{g=0}} \approx e^{-\sigma_p^2g^2} p_0g\,\mathrm{Im}\!\left(\langle\widehat{A}\rangle_w\right).
\label{eq4}
\end{equation}
Here, the relative intensity shift $\Delta L$ is defined from the measured APD voltage $V$ as $\Delta L = ({V - V_{g=0}})/{V_{g=0}}$.

The phase difference measurement accuracy $\delta g$ is given by the following formula: 
\begin{equation}
\delta g = \frac{\delta m}{\left( \partial S \!/\! \partial g \right)},
  \label{eq5}
\end{equation}
where $\delta m$ is the measurement resolution of the P-pointer (determined by the detection of the momentum spectrum distribution) or the intensity uncertainty of the I-pointer. $S$ corresponds to $\Delta p$ and $\Delta l$, and $\partial S \!/\! \partial g$ represents the shift rate.

\begin{figure}[htbp]
\centering
\includegraphics[width=0.48\textwidth]{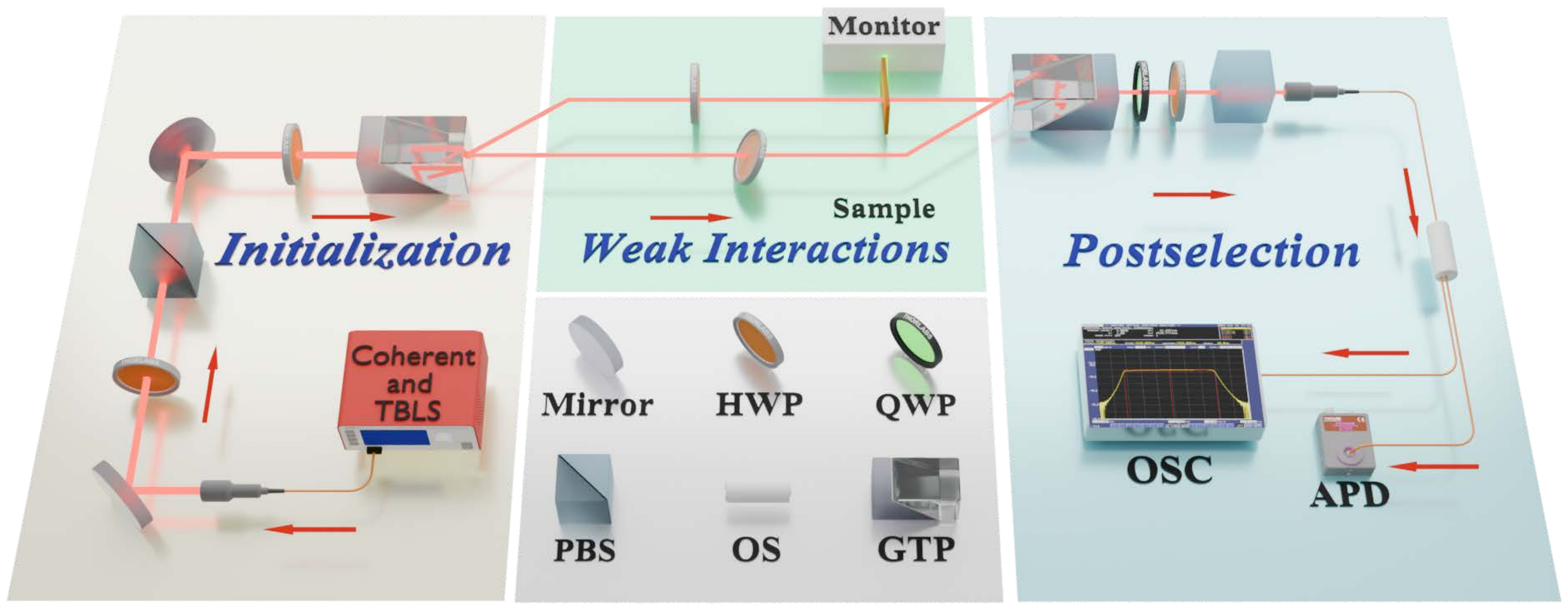}
\caption{\label{fig:1}Experimental schematic of high-sensitivity and high-resolution with weak measurement. TBLS: tunable bandwidth light source; HWP: half-wave plate; PBS: polarization beam splitter; PBD: polarization beam displacer; QWP: quarter-wave plate; GTP: Glan–Taylor polarizer; Membrane: The replacement positions of different thin membrane slide; Monitor: monitoring the angle of the tested slide. The final meter states are separated by a beam splitter into two complementary detections: the spectrometer measures the P-pointer shift while the avalanche photodiode (APD) records the I-pointer shift.}
\end{figure}

The principle of detecting the birefringence coefficient of thin membranes using the shared optical path weak measurement system is shown in Fig.~\ref{fig:1}. A tunable bandwidth light source (TBLS) with flat 1550nm-centered spectrum and a 400 kHz narrow-linewidth coherent source are used. The system uses polarization degree of freedom of the photon as the system observables and momentum spectrum as the measurement instrument state. The pre-selection module combines a HWP and Glan-Taylor prism (GTP) to align partially polarized light's main axis with the GTP transmission axis. Weak interaction occurs through two near-orthogonal HWPs, where one is tilted to create a controlled phase delay. The test sample is placed after this stage, with an auxiliary path monitoring tilt consistency. It is a 2 mm glass slide carries a 5 nm-thick membrane, denoted as $d$ = 5 nm. Post-selection employs a quarter-wave plate (QWP), HWP and GTP combination to generate circular polarization components for imaginary weak value detection. To achieve a high signal-to-noise ratio, we employ the high-stability coherent laser source with a linewidth of 400 kHz and a high-gain avalanche photodiode (APD). Dynamic filtering (a tunable bandwidth filter) and repeated measurements are used to suppress noise impacts. And the wide spectral characteristics of the TBLS ensures high signal-to-noise ratio and accuracy amidst noise. Our spectrometer has a resolution of 0.04 pm.

\begin{figure}[htbp]
\centering
\includegraphics[width=0.48\textwidth]{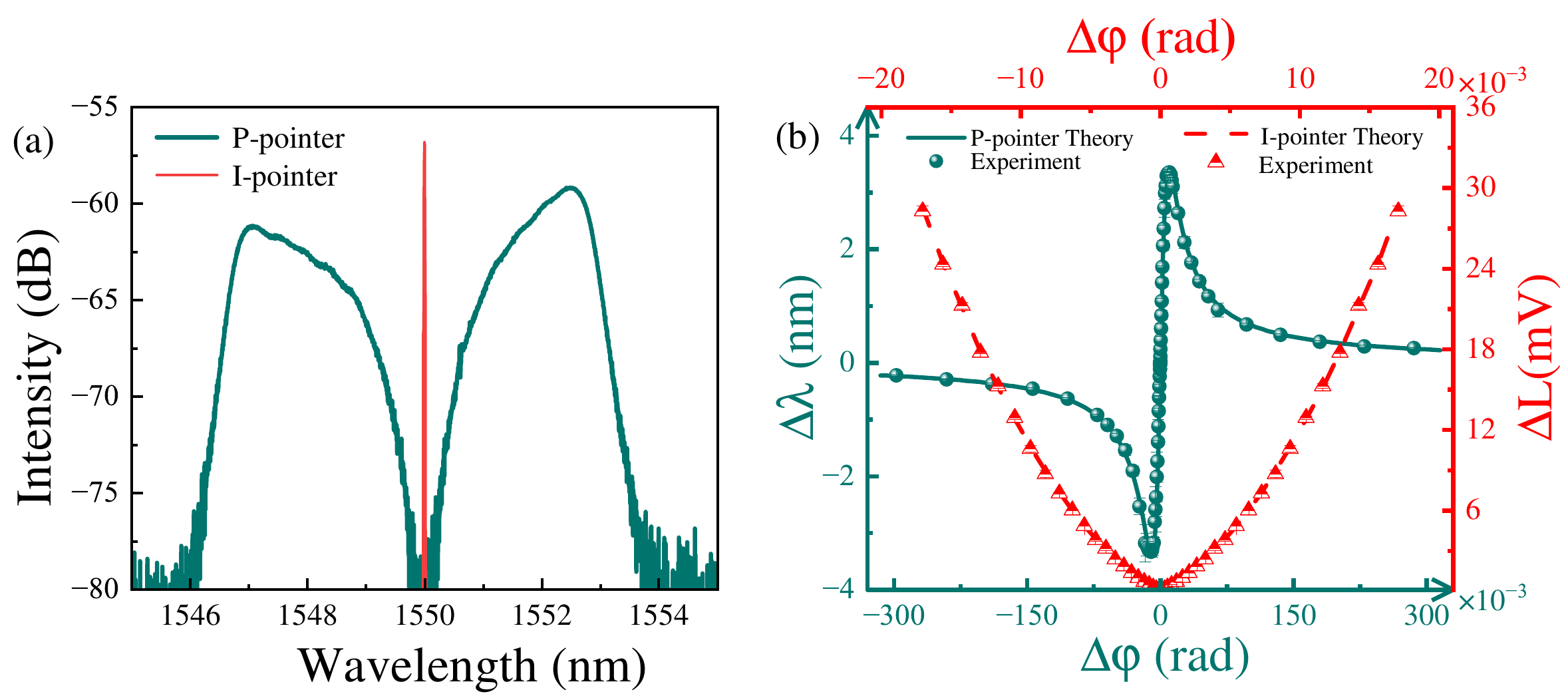}
\caption{\label{fig:2}(a) Measured dual-pointer spectra without phase difference. (b) Intensity shift and spectral shift as functions of phase difference.}
\end{figure}

Figure~\ref{fig:2}(a) shows the spectra measured simultaneously by two pointers with no initial phase difference. Figure~\ref{fig:2}(b) further shows the experimental and theoretical results of the spectral shift and intensity shift as a function of the phase difference. Under the incident light sources, a small phase difference induces significant shifts in both the central wavelength and intensity, clearly revealing the ultra-sensitive response mechanism of weak measurement to rotation angle changes.

\begin{figure}[htbp]
\centering
\includegraphics[width=0.48\textwidth]{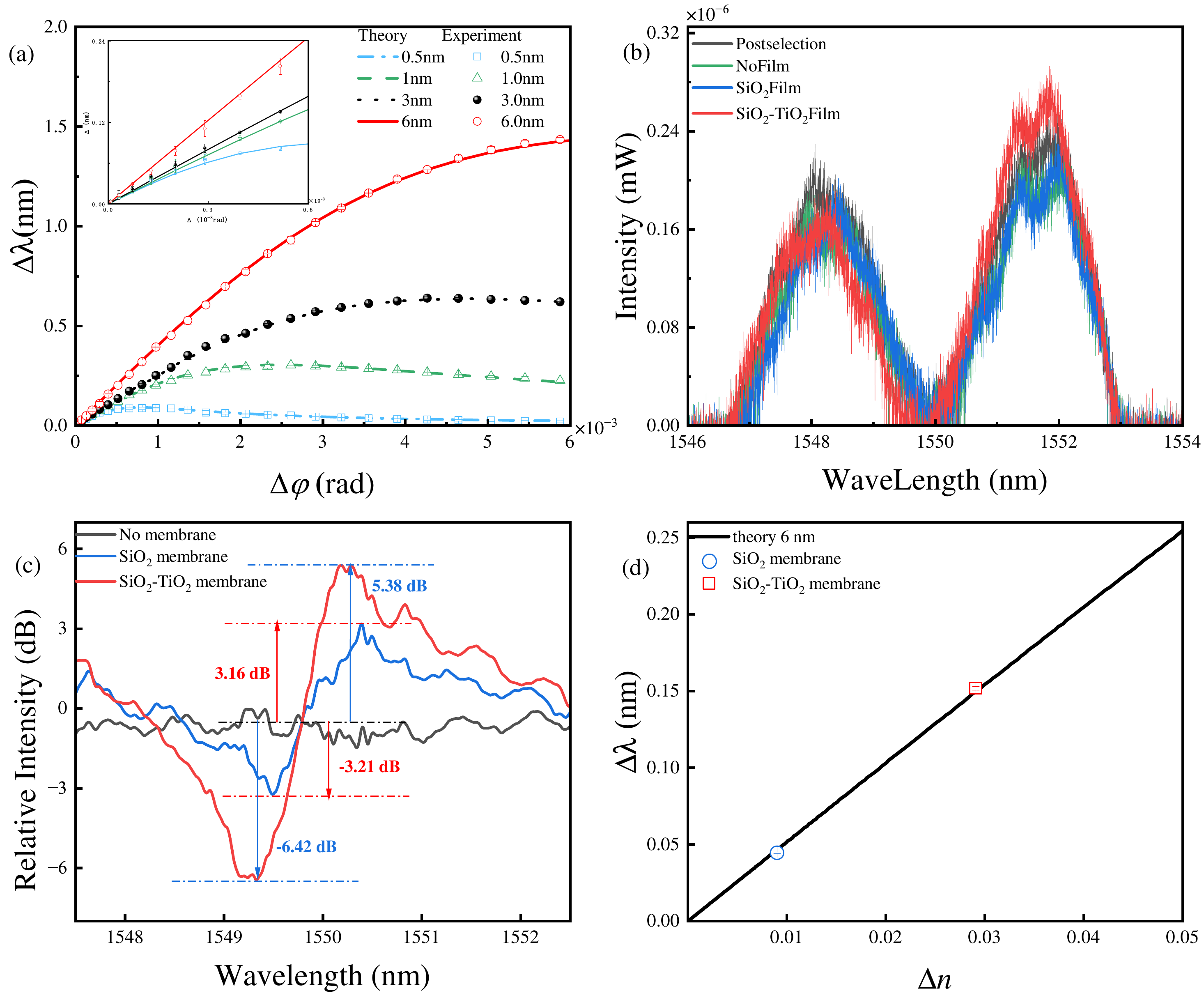}
\caption{\label{fig:3}(a) Results of phase difference versus spectral shift for various spectral widths. (b) Measured spectra. (c) Relative intensity results based on the dual-peak state serving as the reference under a 6 nm broadband light source. (d) Linear fitting of spectral shifts for a spectral width of 6 nm, with error bars representing the standard deviation of 20 measurements.}
\end{figure}

Figure~\ref{fig:3}(a) shows the correspondence between spectral shift and phase difference. At $\rho=0.002$ rad and a system-induced initial phase difference of $\gamma\rho \sim (19/10)\pi$, where $\gamma$ is an initial coupling factor, the optimal agreement between theory and experiment is achieved. From a spectral width of 0.5 nm to 6 nm, as the TBLS spectral width increases, for the same $\Delta \varphi$, higher $\Delta \lambda$ corresponds to higher sensitivity, measurement accuracy and a larger linear range. When $\Delta \varphi$ is in the range of (0, 3$\times 10^{-3}$ rad), the spectral width of 6 nm has the largest linear range. According to Eq.~\ref{eq1}, the range of $\Delta n$ at this point is (0, 0.148). The squares represent the experimental results, while the lines correspond to the theoretical simulation results from Eq.~\ref{eq3}. Using Eq.~\ref{eq5} and the TBLS with spectral widths of 0.5 nm, 1.0 nm, 3.0 nm, and 6.0 nm prepared experimentally, the measured linear phase difference shift rates of $\Delta\lambda/\Delta\varphi$ are 62.0 nm/rad, 190.9 nm/rad, 251.2 nm/rad, and 352.2 nm/rad, respectively. The corresponding phase difference accuracies are 6.45$\times 10^{-7}$ rad, 2.09$\times 10^{-7}$ rad, 1.59$\times 10^{-7}$ rad and 1.14$\times 10^{-7}$ rad, respectively. On this basis, the 6 nm is selected as the optimal bandwidth for further investigation. Notably, the system achieves a high resolution of 1.12$\times 10^{-8}$ RIU and a high sensitivity of 3571 nm/RIU, demonstrating high performance in both sensitivity and resolution. Under weak measurement via post-selection, this transitions from a flat-top profile to a double-peak structure, as shown in Fig.~\ref{fig:3}(b). Figure~\ref{fig:3}(c) shows relative intensity variations under different membranes. Using uncoated glass as a baseline, a $\mathrm{SiO_2}$ membrane induces +3.16 dB and -3.21 dB intensity shifts, corresponding to approximately 50\% signal change, attributed to birefringence differences between the membrane and air. A $\mathrm{SiO_2 - TiO_2}$ membrane produces +5.38 dB and -6.21 dB shifts, exceeding fourfold intensity enhancement, with superimposed birefringence effects amplifying bimodal signals. Subsequent investigation focuses on center wavelength shifts induced by thin membranes with different birefringence coefficients. As shown in Fig.~\ref{fig:3}(d), the birefringent coefficient change induced by a $\mathrm{SiO_2}$ thin membrane slide is measured to be 0.009 RIU, while the birefringent coefficient change induced by a $\mathrm{SiO_2 - TiO_2}$ thin membrane slide is measured to be 0.0291 RIU. Compared to the uncoated slide, these membranes induce center wavelength shifts of 0.0446 nm and 0.1520 nm, respectively.

\begin{figure}[htbp]
\centering
\includegraphics[width=0.48\textwidth]{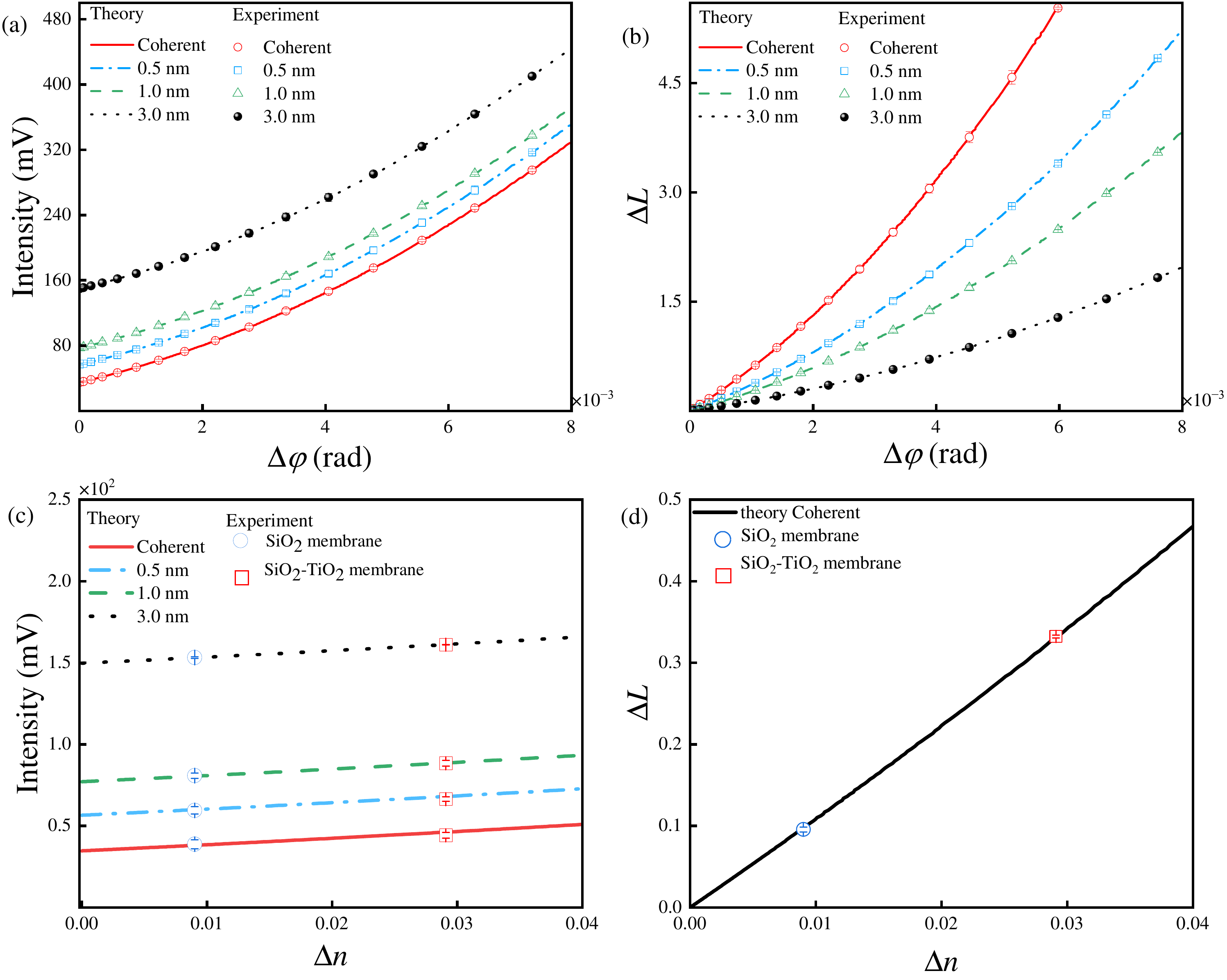}
\caption{\label{fig:4}(a) Total intensity of various spectral width light sources. (b) I-pointer shift $\Delta L$ as a function of phase difference $\Delta \varphi$. The error bars represent the standard deviation of 20 repeated experiments. (c) Linear correspondence between the intensity and the birefringence coefficient under various spectral widths. (d) linear correspondence between the birefringence coefficient and the I-pointer shift.}
\end{figure}

Figure~\ref{fig:4} illustrates the post-selection light intensity and shift $\Delta L$ of the I-pointer as a function of the $\Delta \varphi$. In Fig.~\ref{fig:4}(a), the intensity increases with the spectral width. This trend is attributed to the fact that the broader spectral width results in the higher SNR after post-selection. Figure~\ref{fig:4}(b) shows an inverse relationship between the intensity shift of the I-pointer and the incident spectral bandwidth, exhibiting a contrasting trend to that of the P-pointer. This experimental results agree well with the theory, where under narrow-spectrum illumination, the meter state demonstrates an evident longitudinal shift but minimal transverse shift, corresponding to a noticeable I-pointer shift and negligible P-pointer shift. According to Eq.~\ref{eq5}, the intensity departure for coherent light and TBLS with spectral widths of 0.5 nm, 1.0 nm, 3.0 nm are 0.044 mV, 0.072 mV, 0.11 mV, and 0.21 mV, respectively, and the corresponding phase difference accuracy $\delta g$ is 2.02$\times 10^{-6}$ rad, 3.17$\times 10^{-6}$ rad, 4.83$\times 10^{-6}$ rad and 9.37$\times 10^{-6}$ rad. On this basis, the coherent laser is chosen as the optimal illumination for further investigation, notably achieving a high sensitivity of 4710 mV/RIU and a high resolution of 9.34$\times 10^{-6}$ RIU. The sensitivity is above 1500 nm/RIU higher than that of existing surface plasmon resonance detection, and the resolution is sufficient to cover the range of birefringence coefficients for typical crystals. For the intensity changes and I-pointer shifts caused by different birefringent coefficient thin membrane slides in the above experiment, further processing and analysis are carried out. As depicted in Fig.~\ref{fig:4}(c), to optimize the balance between the largest linear region and the lowest SNR, we select a coherent laser with a linewidth of 400 kHz as the main measurement. In Fig.~\ref{fig:4}(d), the birefringent coefficient change caused by a $\mathrm{SiO_2}$ membrane and a $\mathrm{SiO_2 - TiO_2}$ membrane is 0.009 RIU and 0.0291RIU, respectively. This result is consistent with the P-pointer measurement data, further validating the feasibility of measurements using both the P and I pointers. Relative to the uncoated slide, the $\mathrm{SiO_2}$ membrane causes an I-pointer shift of 0.0955, while the $\mathrm{SiO_2 - TiO_2}$ membrane shifts it by 0.3322.

In conclusion, this work introduces and experimentally validates a high-sensitivity and high-resolution weak measurement system for nanoscale membrane birefringence coefficient determination. By simultaneously employing tunable-bandwidth light and coherent light sources, the complementary advantages of variable bandwidth are fully demonstrated. The tunable bandwidth light optimizes the dynamic range and noise immunity of the P-pointer, while the coherent light enhances the sensitivity of the I-pointer, all achieved calibration-free operation across various bandwidths. When the spectral width is 6 nm, the system achieves the best resolution of 1.12 $\times 10^{-8}$ RIU, significantly outperforming traditional ellipsometers. The optimal sensitivity is achieved with the narrow-linewidth coherent laser, reaching 4710 mV/RIU, which surpasses existing surface plasmon resonance measurements by above 1500 mV/RIU. The system's linear range is from 0 to 0.148, fully covering the birefringence coefficient range of the typical crystals. Moreover, the setup of the auxiliary optical path eliminates substrate interference, thereby extending the detection limit to $10^{-8}$ RIU. Our work develops a high-precision and robust solution for optical quantum measurement of advanced nonmaterial in challenging environments.

The authors thank Professor Guohui Li for his valuable assistance during the membrane preparation stage. This work was supported by the National Key Research and Development Program of China (2022YFA1404201), the National Natural Science Foundation of China (62475185, 62175176, U23A20380), and the Fundamental Research Program of Shanxi Province (202403021221034).

\section*{AUTHOR DECLARATIONS}
\vspace{-0.15in}
\subsection*{Conflict of Interest}
\vspace{-0.15in}
The authors have no conflicts to disclose.

\subsection*{Author Contributions}
\vspace{-0.15in}
\noindent \textbf{Shuqi Gao}: Investigation (equal); Formal analysis (equal); Data curation (equal); Methodology (equal); Software (equal); Validation (equal); Writing – original draft (equal). 
\textbf{Min Zhang}: Data curation (equal); Investigation (supporting); Methodology (supporting). 
\textbf{Jiahui Hou}: Investigation (supporting); Software (supporting). 
\textbf{Qingchen Liu}: Investigation (supporting). 
\textbf{Hongyu Li}: Investigation (supporting); Validation (equal). 
\textbf{Xiaomin Guo}: Conceptualization (equal); Investigation (lead); Funding acquisition (equal); Project administration (equal); Supervision (equal); Writing – review $\&$ editing (equal).
\textbf{Yanqiang Guo}: Conceptualization (equal); Formal analysis (equal); Data curation (equal); Investigation (lead); Methodology (lead); Funding acquisition (equal); Project administration (equal); Software (equal); Supervision (lead); Validation (equal); Writing – original draft (lead);  Writing – review$\&$ editing (lead).
\textbf{Liantuan Xiao}: Conceptualization (equal); Investigation (lead); Funding acquisition (equal); Project administration (equal); Supervision (equal); Writing – review $\&$ editing (equal).

\section*{DATA AVAILABILITY}
\vspace{-0.15in}
The data that support the findings of this study are available from the corresponding authors upon reasonable request.

\section*{REFERENCES}
\vspace{-0.2in}
\bibliography{ref}% 

@PREAMBLE{
 "\providecommand{\noopsort}[1]{}" 
 # "\providecommand{\singleletter}[1]{#1}%" 
}

@article{Jia25,
  title = {The birefringence in orthorhombic and triclinic lead-iodates and the functional basic units' contribution: A first-principles investigation},
  journal = {Chemical Physics Letters},
  volume = {867},
  pages = {141987},
  year = {2025},
  issn = {0009-2614},
  doi = {https://doi.org/10.1016/j.cplett.2025.141987},
  url = {https://www.sciencedirect.com/science/article/pii/S0009261425001277},
  author = {Jialong Wang and Enshen Wang and Xiuhua Cui and Peng Li and Qun Jing and Zhaohui Chen}
}

@article{Mas25,
  title = {Inhomogeneous birefringence analysis using a tensor-valued backprojection},
  author = {Masafumi Seigo; Hidetoshi Fukui; Shogo Kawano; Meredith Kupinski},
  journal = {Optical Review},
  volume = {59},
  issue = {3},
  pages = {278--281},
  numpages = {0},
  year = {2025},
  month = {Mar},
  publisher = {Optical Review},
  doi = {10.1021/ACS.INORGCHEM.5C00306},
  url = {https://doi.org/10.1007/s10043-025-00954-3}
}

@article{Xav24,
author = {Xavier Theillier and Sylvain Rivet and Matthieu Dubreuil and Yann Le Grand},
journal = {Opt. Lett.},
number = {2},
pages = {387--390},
publisher = {Optica Publishing Group},
title = {Swept-wavelength null polarimetry for highly sensitive birefringence laser scanning microscopy},
volume = {49},
month = {Jan},
year = {2024},
url = {https://opg.optica.org/ol/abstract.cfm?URI=ol-49-2-387},
doi = {10.1364/OL.507576}
}

@article{Kur25,
    author = "Kuratsuji, Hiroshi",
    title = "{Optical skyrmions in nonlinear birefringence media: semiclassical approach incorporating effective gauge field}",
    doi = "10.1088/2040-8986/ad9ab4",
    journal = "J. Opt. A",
    volume = "27",
    number = "1",
    pages = "015501",
    year = "2025"
}

@article{Das24,
title = {Achromatic correction for birefringent interferometers that improve Fourier transform spectrometers and hyperspectral imaging},
author={Dasol Im and Zachary M Faitz and Feng Jin and Joo Soo Kim and Erica Magee and Priyanthi Amarasinghe and Sudhir Trivedi and Martin T Zanni},
journal={Optical Express},
year={2024},
doi = {10.1364/OE.538565},
url = {https://pubmed.ncbi.nlm.nih.gov/39573760/},
}

@article{Sun25,
title = {Large-format grating groove density measurement method based on optical interferometry},
journal = {Optics and Lasers in Engineering},
volume = {187},
pages = {108885},
year = {2025},
issn = {0143-8166},
doi = {https://doi.org/10.1016/j.optlaseng.2025.108885},
url = {https://www.sciencedirect.com/science/article/pii/S0143816625000727},
author = {Yujia Sun and Wenyuan Zhou and Zhaowu Liu and Wenhao Li and Shan Jiang and Lin Liu and Yanxiu Jiang and Weicheng Wang},
}

@article{KIM22,
title = {Surface measurement of silicon wafer using harmonic phase-iterative analysis and wavelength-scanning Fizeau interferometer},
journal = {Precision Engineering},
volume = {75},
pages = {142-152},
year = {2022},
issn = {0141-6359},
doi = {https://doi.org/10.1016/j.precisioneng.2022.02.005},
url = {https://www.sciencedirect.com/science/article/pii/S0141635922000344},
author = {Sungtae Kim and Yangjin Kim and Naohiko Sugita and Mamoru Mitsuishi},
}

@article{Jo07,
  title = {Complex weak values in quantum measurement},
  author = {Jozsa, Richard},
  journal = {Phys. Rev. A},
  volume = {76},
  issue = {4},
  pages = {044103},
  numpages = {3},
  year = {2007},
  month = {Oct},
  publisher = {American Physical Society},
  doi = {10.1103/PhysRevA.76.044103},
  url = {https://link.aps.org/doi/10.1103/PhysRevA.76.044103}
}

@article{You18,
  title = {Phase amplification in optical interferometry with weak measurement},
  author = {Li, Li and Li, Yuan and Zhang, You-Lang and Yu, Sixia and Lu, Chao-Yang and Liu, Nai-Le and Zhang, Jun and Pan, Jian-Wei},
  journal = {Phys. Rev. A},
  volume = {97},
  issue = {3},
  pages = {033851},
  numpages = {6},
  year = {2018},
  month = {Mar},
  publisher = {American Physical Society},
  doi = {10.1103/PhysRevA.97.033851},
  url = {https://link.aps.org/doi/10.1103/PhysRevA.97.033851}
}

@article{JS11,
author = {Jeff S. Lundeen and Brandon Sutherland and Aabid Patel and Corey Stewart and Charles Bamber},
journal = {Nature},
number = {6},
pages = {8461--8473},
publisher = {Optica Publishing Group},
title = {Direct measurement of the quantum wavefunction},
volume = {30},
month = {Jun},
year = {2011},
url = {https://doi.org/10.1038/nature10120},
doi = {10.1038/nature10120},
}

@article{Fe14,
  title = {Weak Value Amplification is Suboptimal for Estimation and Detection},
  author = {Ferrie, Christopher and Combes, Joshua},
  journal = {Phys. Rev. Lett.},
  volume = {112},
  issue = {4},
  pages = {040406},
  numpages = {5},
  year = {2014},
  month = {Jan},
  publisher = {American Physical Society},
  doi = {10.1103/PhysRevLett.112.040406},
  url = {https://link.aps.org/doi/10.1103/PhysRevLett.112.040406}
}

@article{Q17,
    author = {Qiu, Xiaodong and Xie, Linguo and Liu, Xiong and Luo, Lan and Li, Zhaoxue and Zhang, Zhiyou and Du, Jinglei},
    title = {Precision phase estimation based on weak-value amplification},
    journal = {Applied Physics Letters},
    volume = {110},
    number = {7},
    pages = {071105},
    year = {2017},
    month = {02},
    doi = {10.1063/1.4976312},
    url = {https://doi.org/10.1063/1.4976312},
}

@article{Xu20,
  title = {Approaching Quantum-Limited Metrology with Imperfect Detectors by Using Weak-Value Amplification},
  author = {Xu, Liang and Liu, Zexuan and Datta, Animesh and Knee, George C. and Lundeen, Jeff S. and Lu, Yan-qing and Zhang, Lijian},
  journal = {Phys. Rev. Lett.},
  volume = {125},
  issue = {8},
  pages = {080501},
  numpages = {6},
  year = {2020},
  month = {Aug},
  publisher = {American Physical Society},
  doi = {10.1103/PhysRevLett.125.080501},
  url = {https://link.aps.org/doi/10.1103/PhysRevLett.125.080501}
}

@article{Luo19,
    author = {Luo, Lan and Xie, Linguo and Qiu, Jiangdong and Zhou, Xinxing and Liu, Xiong and Li, Zhaoxue and He, Yu and Zhang, Zhiyou and Sun, Handong},
    title = {Simultaneously precise estimations of phase and amplitude variations based on weak-value amplification},
    journal = {Applied Physics Letters},
    volume = {114},
    number = {11},
    pages = {111104},
    year = {2019},
    month = {03},
    doi = {10.1063/1.5083995},
    url = {https://doi.org/10.1063/1.5083995},
}

@article{Peng,
  title = {Experimental super-Heisenberg quantum metrology with indefinite gate order},
  author = {Peng Yin and Xiaobin Zhao and Yuxiang Yang and Yu Guo and Wen-Hao Zhang and Gong-Chu Li and Yong-Jian Han and Bi-Heng Liu and Jin-Shi Xu and Giulio Chiribella and Geng Chen and Chuan-Feng Li & Guang-Can Guo },
   year={2023},
  journal = {Nature Physics},
  month = {01},
 volume = {19},
   pages = {1122–1127 }, 
  url = {https://doi.org/10.1038/s41567-023-02046-y}
}

@article{Bru10,
  title = {Measuring Small Longitudinal Phase Shifts: Weak Measurements or Standard Interferometry?},
  author = {Brunner, Nicolas and Simon, Christoph},
  journal = {Phys. Rev. Lett.},
  volume = {105},
  issue = {1},
  pages = {010405},
  numpages = {4},
  year = {2010},
  month = {Jul},
  publisher = {American Physical Society},
  doi = {10.1103/PhysRevLett.105.010405},
  url = {https://link.aps.org/doi/10.1103/PhysRevLett.105.010405}
}

@article{Xu13,
  title = {Phase Estimation with Weak Measurement Using a White Light Source},
  author = {Xu, Xiao-Ye and Kedem, Yaron and Sun, Kai and Vaidman, Lev and Li, Chuan-Feng and Guo, Guang-Can},
  journal = {Phys. Rev. Lett.},
  volume = {111},
  issue = {3},
  pages = {033604},
  numpages = {4},
  year = {2013},
  month = {Jul},
  publisher = {American Physical Society},
  doi = {10.1103/PhysRevLett.111.033604},
  url = {https://link.aps.org/doi/10.1103/PhysRevLett.111.033604}
}

@article{Hu19,
  title = {Toward ultrahigh sensitivity in weak-value amplification},
  author = {Huang, Jingzheng and Li, Yanjia and Fang, Chen and Li, Hongjing and Zeng, Guihua},
  journal = {Phys. Rev. A},
  volume = {100},
  issue = {1},
  pages = {012109},
  numpages = {6},
  year = {2019},
  month = {Jul},
  publisher = {American Physical Society},
  doi = {10.1103/PhysRevA.100.012109},
  url = {https://link.aps.org/doi/10.1103/PhysRevA.100.012109}
}

@article{luo1,
  title = {Spatial differential operation and edge detection based on the geometric spin Hall effect of light},
  journal = {Optics Letters},
  volume = {45},
  number = {4},
  pages = {877--880},
  year = {2020},
  issn = {0146-9592},
  doi = {10.1364/OL.386224},
  url = {https://opg.optica.org/ol/abstract.cfm?uri=ol-45-4-877},
  author = {Shanshan He and Junxiao Zhou and Shizhen Chen and Weixing Shu and Hailu Luo and Shuangchun Wen}
}

@article{luo2,
  title = {Large in-plane asymmetric spin angular shifts of a light beam near the critical angle},
  journal = {Optics Letters},
  volume = {44},
  number = {2},
  pages = {207--210},
  year = {2019},
  issn = {0146-9592},
  doi = {10.1364/OL.44.000207},
  url = {https://opg.optica.org/ol/abstract.cfm?uri=ol-44-2-207},
  author = {Xinxing Zhou and Linguo Xie and Xiaohui Ling and Shijia Cheng and Zhiyou Zhang and Hailu Luo and Handong Sun}
}

@article{luo3,
  title = {Photonic spin Hall effect in dielectric metasurfaces with rotational symmetry breaking},
  journal = {Optics Letters},
  volume = {40},
  number = {5},
  pages = {756--759},
  year = {2015},
  issn = {0146-9592},
  doi = {10.1364/OL.40.000756},
  url = {https://opg.optica.org/ol/abstract.cfm?uri=ol-40-5-756},
  author = {Yachao Liu and Xiaohui Ling and Xunong Yi and Xinxing Zhou and Shizhen Chen and Yougang Ke and Hailu Luo and Shuangchun Wen}
}

@article{zhang1,
  title = {Optical differentiation based on weak measurements},
  journal = {Optics Letters},
  volume = {47},
  number = {15},
  pages = {3880--3883},
  year = {2022},
  issn = {0146-9592},
  doi = {10.1364/OL.463016},
  url = {https://opg.optica.org/ol/abstract.cfm?uri=ol-47-15-3880},
  author = {An Wang and Junfan Zhu and Lan Luo and Xiong Liu and Ling Ye and Zhiyou Zhang and Jinglei Du}
}

@article{zhang2,
  title = {Anomalous amplification in almost-balanced weak measurement for measuring spin Hall effect of light},
  journal = {Optics Express},
  volume = {28},
  number = {5},
  pages = {6408--6416},
  year = {2020},
  issn = {1094-4087},
  doi = {10.1364/OE.386017},
  url = {https://opg.optica.org/oe/abstract.cfm?uri=oe-28-5-6408},
  author = {Lan Luo and Yu He and Xiong Liu and Zhaoxue Li and Pi Duan and Zhiyou Zhang}
}

@article{Li11,
  title = {Ultrasensitive phase estimation with white light},
  author = {Li, Chuan-Feng and Xu, Xiao-Ye and Tang, Jian-Shun and Xu, Jin-Shi and Guo, Guang-Can},
  journal = {Phys. Rev. A},
  volume = {83},
  issue = {4},
  pages = {044102},
  numpages = {4},
  year = {2011},
  month = {Apr},
  publisher = {American Physical Society},
  doi = {10.1103/PhysRevA.83.044102},
  url = {https://link.aps.org/doi/10.1103/PhysRevA.83.044102}
}

@article{zhang3,
  title = {Estimation of optical rotation of chiral molecules with weak measurements},
  journal = {Optics Letters},
  volume = {41},
  number = {17},
  pages = {4032--4035},
  year = {2016},
  issn = {0146-9592},
  doi = {10.1364/OL.41.004032},
  url = {https://opg.optica.org/ol/abstract.cfm?uri=ol-41-17-4032},
  author = {Xiaodong Qiu and Linguo Xie and Xiong Liu and Lan Luo and Zhiyou Zhang and Jinglei Du}
}

@article{zhang4,
  title = {Measurement of hysteresis loop based on weak measurement},
  journal = {Optics Letters},
  volume = {45},
  number = {5},
  pages = {1075--1078},
  year = {2020},
  issn = {0146-9592},
  doi = {10.1364/OL.383764},
  url = {https://opg.optica.org/ol/abstract.cfm?uri=ol-45-5-1075},
  author = {Qi Wang and Tong Li and Lan Luo and Yu He and Xiong Liu and Zhaoxue Li and Zhiyou Zhang and Jinglei Du}
}

@article{zhang5,
  title = {Estimation of Kerr angle based on weak measurement with two pointers},
  journal = {Optics Express},
  volume = {31},
  number = {9},
  pages = {14432--14441},
  year = {2023},
  issn = {1094-4087},
  doi = {10.1364/OE.487363},
  url = {https://opg.optica.org/oe/abstract.cfm?uri=oe-31-9-14432},
  author = {Lan Luo and Tong Li and Yinghang Jiang and Liang Fang and Bo Liu and Zhiyou Zhang}
}

@article{zhang6,
  title = {Detection of magneto-optical Kerr signals via weak measurement with frequency pointer},
  journal = {Optics Letters},
  volume = {46},
  number = {17},
  pages = {4140--4143},
  year = {2021},
  issn = {0146-9592},
  doi = {10.1364/OL.428486},
  url = {https://opg.optica.org/ol/abstract.cfm?uri=ol-46-17-4140},
  author = {Yu He and Lan Luo and Linguo Xie and Jingyi Shao and Yurong Liu and Jiacheng You and Yucheng Ye and Zhiyou Zhang}
}

@article{he1,
  title = {Optimization of a quantum weak measurement system with digital filtering technology},
  journal = {Applied Optics},
  volume = {57},
  number = {27},
  pages = {7956--7966},
  year = {2018},
  issn = {1559-128X},
  doi = {10.1364/AO.57.007956},
  url = {https://opg.optica.org/ao/abstract.cfm?uri=ao-57-27-7956},
  author = {Yang Xu and Lixuan Shi and Tian Guan and Dongmei Li and Yuxuan Yang and Xiangnan Wang and Zhangyan Li and Luyuan Xie and Xuesi Zhou and Yonghong He and Wenyue Xie}
}

@article{he2,
  title = {Optimization of a quantum weak measurement system with its working areas},
  journal = {Optics Express},
  volume = {26},
  number = {16},
  pages = {21119--21131},
  year = {2018},
  issn = {1094-4087},
  doi = {10.1364/OE.26.021119},
  url = {https://opg.optica.org/oe/abstract.cfm?uri=oe-26-16-21119},
  author = {Yang Xu and Lixuan Shi and Tian Guan and Cuixia Guo and Dongmei Li and Yuxuan Yang and Xiangnan Wang and Luyuan Xie and Yonghong He and Wenyue Xie}
}

@article{he3,
  title = {In situ detection of electrochemical reaction by weak measurement},
  journal = {Optics Express},
  volume = {29},
  number = {13},
  pages = {19292--19304},
  year = {2021},
  issn = {1094-4087},
  doi = {10.1364/OE.426345},
  url = {https://opg.optica.org/oe/abstract.cfm?uri=oe-29-13-19292},
  author = {Zhangyan Li and Yang Xu and Kaijie Ma and Le Liu and Jingyu Xi and Tian Guan and Fuying Li and Chongqi Zhou and Suyi Zhong and Yonghong He}
}

@article{guo1,
  title = {Detecting momentum weak value: {Shack--Hartmann} versus a weak measurement wavefront sensor},
  journal = {Optics Letters},
  volume = {46},
  number = {21},
  pages = {5352--5355},
  year = {2021},
  issn = {0146-9592},
  doi = {10.1364/OL.439174},
  url = {https://opg.optica.org/ol/abstract.cfm?uri=ol-46-21-5352},
  author = {Yi Zheng and Mu Yang and Zheng-Hao Liu and Jin-Shi Xu and Chuan-Feng Li and Guang-Can Guo}
}

@article{guo2,
  title = {Experimental realization of sequential weak measurements of non-commuting Pauli observables},
  journal = {Optics Express},
  volume = {27},
  number = {5},
  pages = {6089--6097},
  year = {2019},
  issn = {1094-4087},
  doi = {10.1364/OE.27.006089},
  url = {https://opg.optica.org/oe/abstract.cfm?uri=oe-27-5-6089},
  author = {Jiang-Shan Chen and Meng-Jun Hu and Xiao-Min Hu and Bi-Heng Liu and Yun-Feng Huang and Chuan-Feng Li and Can-Guang Guo and Yong-Sheng Zhang}
}

@misc{guo2024,
    author = "Guo, Yanqiang and Zhang, Jianchao and Hou, Jiahui and Guo, Xiaomin and Xiao, Liantuan",
    title = "{Ultraprecise time-difference measurement via enhanced dual pointers with multiple weak interactions}",
    eprint = "2405.06863",
    archivePrefix = "arXiv",
    primaryClass = "quant-ph",
    month = "5",
    year = "2024"
}
\end{document}